# Tailored business solutions by workflow technologies

**Assist.Prof.Florin Fortiş, PhD, Prof. Alexandru Cicortaş, PhD,**
The West Universiy of Timişoara, România
**Assist.Prof. Alexandra Fortiş, PhD Candidate,**
The "Tibiscus" University of Timişoara, România

REZUMAT. VISP (Virtual Internet Service Provider) is an IST-STREP project, which is conducting research in the field of these new technologies, targeted to telecom/ISP companies. One of the first tasks of the VISP project is to identify the most appropriate technologies in order to construct the VISP platform. This paper presents the most significant results in the field of choreography and orchestration, two key domains that must accompany process modeling in the construction of a workflow environment.

## 1. Introduction

VISP (Virtual Internet Service Provider) is an IST-STREP project aimed to providing a software platform to enable small and medium-sized enterprises (SMEs) to cooperate and offer an environment such that they can operate as a single business (a business cluster, or "IST cluster"). Thus, they are able to provide tailored, targeted solutions, adapted to the needs of different businesses. However, the collaboration between the partners of a cluster could be effective only if most of the activities can be completely automated.

By using a VISP cluster, partners are able to offer specific services (several niche, value-added services) in addition to their basic offer. A VISP cluster can, thus, enable competition between larger telecom companies and SMEs that are participating in a "IST cluster", by enlarging the portfolio of the participating SMEs with services tailored to client's needs, possibly composed of several service building blocks (understood as very basic services for the target industry), provided by different members of the VISP cluster.

Several organizations and standards bodies are highly involved in the continuous transformation and improvement of the field of workflow technologies. Web-services, and web-based or XML-based technologies, together with the continuous evolution of the Internet have changed the entire electronic business environment. Old EDI-based solutions become obsolete, and XML and web-services based solutions are being considered for the creation and usage of these new, improved business environments.

## 2. Target stack of technologies

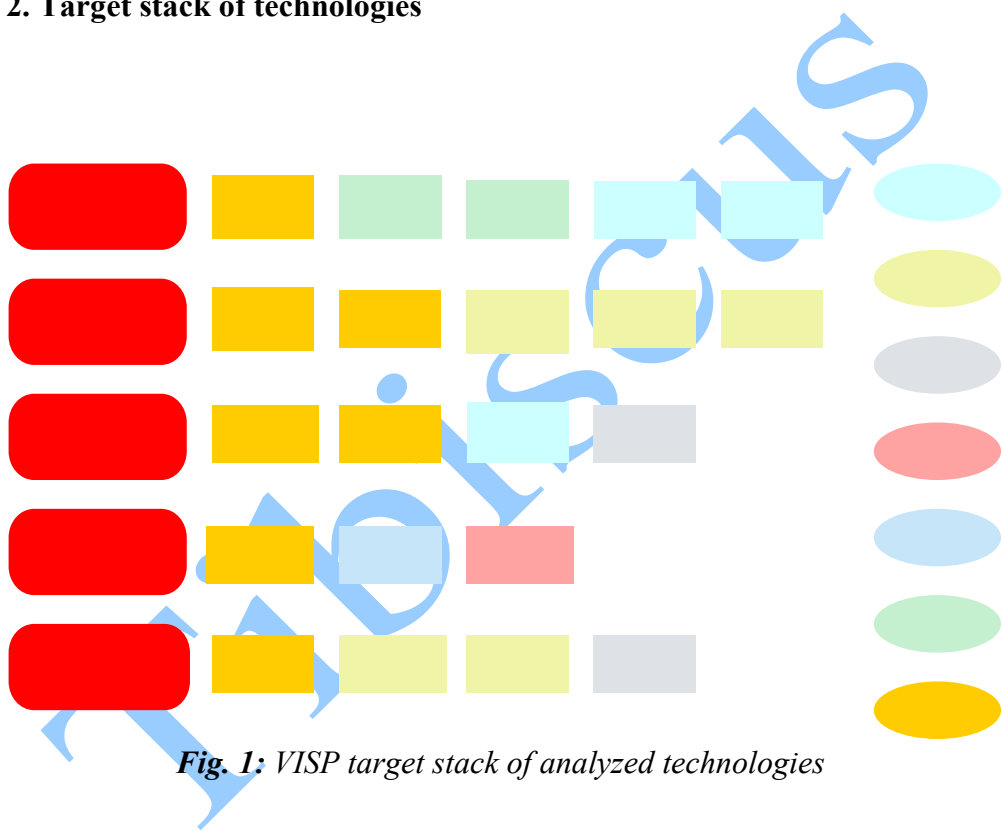

*Fig. 1:* VISP target stack of analyzed technologies

### 2.1 Standards bodies

In our days, there are a large number of organizations that are responsible with the development, adoption, and maintenance of e-business standards. However, the most promising technologies tend to concentrate around several organizations and standards bodies. The VISP project has considered the activity of the most promising technologies from these

Anale. Seria Informatică. Vol. IV fasc. I - 2006
Annals. Computer Science Series. 4th Tome 1st Fasc. - 2006

organizations and standards bodies, technologies targeted to process modeling, orchestration, choreography, and interoperability (communication, control, data layers, and others).

*Table 1:* Standards bodies providing technologies for the VISP project

| Name | Description |
|---|---|
| OASIS (Organization for the Advancement of Structured Information Standards) | A not-for-profit, international consortium that drives the development, convergence, and adoption of e-business standards. |
| OMG (Object Management Group) | An open membership, not-for-profit consortium that produces and maintains computer industry specifications for interoperable enterprise applications. |
| BPMI (Business Process Management Initiative) | A non-profit corporation aimed to promote and develop open standards, XML-based, in order to support the development and operation of business processes spanning multiple applications and business partners over the Internet. |
| WfMC (Workflow Management Coalition) | A non-profit, international organization of workflow vendors, users, analysts and university/research groups. The mission of WfMC is to promote and develop the use of workflow through the establishment of standards for software terminology, interoperability and connectivity between workflow products. |
| W3C (World Wide Web Consortium) | An international consortium developing Web standards, and interoperable technologies (tools, specifications, software). W3C's mission is to lead the World Wide Web to its full potential by developing protocols and guidelines that ensure long-term growth for the Web. |
| RosettaNet | A non-profit organization dedicated to the collaborative development and rapid deployment of open, e-business process standards that align processes within global trading networks. |
| OAGi (Open Applications Group, Inc.) | A not-for-profit open standards group building process-based XML standards for both B2B and A2A integration. |

## 3. Choreography technologies

A choreography defines (as in 0) "*the sequence and conditions under which multiple cooperating independent agents exchange messages in order to perform a task to achieve a goal state.*"

The choreography is focused on the composition of services. If a standard is related to choreography, it has to specify how to compose existing services, the protocols to be considered between different parties either during its normal execution, or in erroneous situations.

53

Choreography languages concentrate on the specification of observable behavior of communicating services. They offer a description of the messages to be exchanged between services, and not a recommendation on how the services have to be implemented in order to be able to exchange the messages. Choreography languages are not executable, but there must be some mappings and transformation between choreographies and orchestrations.

The VISP project requires not only to identify appropriate choreography languages, but also to identify if there are available tools supporting the mapping between choreographies and orchestrations. Several key criteria have been considered for choreography languages in order to make a selection of the most appropriate technologies. These criteria and the overall evaluation for the analyzed standards have been synthesized in Table 2.

*Table 2: Core technical criteria for choreography languages*

| Technical criteria | WS-CDL | BPSS |
|---|---|---|
| Specification of control flow/observable behaviour | + | + |
| Consideration of multiple components; end-to-end behaviour | + | + |
| Complex structures | + | + |
| Specification of data flows | + | + |
| Compatibility with WSDL/BPMN | +/- | +/0 |
| Supports document exchange communication style | + | + |
| Supports RPC communication style | + | + |
| Supports error handling/compensation | +/+ | +/- |
| Supports transaction processing | + | + |
| Support for activity and event-based modeling | + | + |
| Formal definition/Theoretical basis | XSD/ $\pi$—calculus | XSD,CCTS/n.a |
| Extensibility | + | 0 |

## 3.1. BPMN (Business Process Modeling Notation)

BPMN is a standard developed by BPMI. Currently, the development process for this standard has been transferred to OMG, after the merger between the two organizations. BPMN is more a graphical notation than a choreography language. However, its powerful facilities allow its usage both as for process modeling and choreography.

BPMN is a graphical notation. It allows a high-level business process design, provides support for modeling quite complex business processes, and offers the mapping to an executable language, like BPEL or XPDL. However, BPMN is independent of any specific business process modeling methodology.

BPMN offers good mapping to executable languages (such as BPEL, XPDL), and a good support for B2B process concepts (for example, choreography-type interactions between business entities). Unlike other business process notations, BPMN enables data objects to be represented, modelers being ale to show how data, documents and other objects are used and updated during a process flow. However, BPMN is not a data flow diagram notation: data and information models are not included in the specification.

BPMN was designed specifically for business process modeling. Even if its theoretical model is the π-calculus, BPMN maps very well on the workflow patterns, developed by van der Aalst et al. 0. As a graphical notation, BPMN was recommended for use in the VISP project in early stages of system design, particularly for business process flow aspects of the VISP project. The choreography of these business processes can be specified using BPMN notation.

## 3.2. ebXML BPSS (ebXML Business Process Specification Schema)

BPSS provides the semantics, elements, and properties necessary to define *business collaborations*. *Business collaboration* consists of a set of roles collaborating through a set of choreographed *transactions* by exchanging *business documents*. BPSS uses concepts like "*Business Collaboration*", "*Business Transaction*", "*Business document flow*", and "*Choreography*".

ebXML BPSS is an approved OASIS standard, part of the ebXML suite of specifications. Most of the standards from the ebXML suite are approved as ISO 15000 standards. However, the ebXML project has started as a joint effort from OASIS and UN/CEFACT standard bodies, its initial rationale being to provide five layers of specifications, in order to cover the fields of *business processes*, *core data components*, *collaboration protocol agreements*, *messaging*, and *registries and repositories*. Thus, BPSS provides support for collaborative processes by means of CPP/CPA (Collaborative Protocol Profile, and Agreement), providing a generic framework for business process collaborations, both between two parties and multiparty (see also 0).

BPSS is a mature standard, well covered by industry adoption (for more information, see 0). It is highly interoperable with other standards and technologies considered for the VISP project, including the ebXML stack of standards. Thus, BPSS proves to be a good choreography candidate. However, its adoption should be made with certain care, since there is the general impression that no significant progress has been made in the last years for the ebXML stack of standards (this is a false impression, however, considering the recent approval as ISO 15000 standards for five of the ebXML standards).

### 3.3. WS-CDL (Web Services Choreography Description Language)

WS-CDL is, at the moment, a Candidate Recommendation (as of May 2006) at W3C. It is described as being "*an XML-based language that describes peer-to-peer collaborations of participants by defining, from a global viewpoint, their common and complementary observable behavior [...] The WS-CDL specification is targeted for composing interoperable, peer-to-peer collaborations between any type of participant regardless of the supporting platform or programming model used by the implementation of the hosting environment.*" (see 0)

The Web Services Choreography Working Group specifies that the choreography language specification shall define at least the behavior and language constructs in order to cover: *composition features*, *associations*, *message exchanges*, and *state management*. WS-CDL is highly related with other technologies from W3C, including WSDL 2.0, WSCI and WSCL, technologies that were also considered for VISP evaluation (see 0). However, even if the former two standards were analyzed for the VISP project, they are superseded by the WS-CDL, and thus they were not considered as choreography candidates.

WS-CDL is a powerful choreography language that allows choreography specification in great detail. Choreographies are specified in a prescriptive way. Thus, it is possible a derivation of the corresponding executable orchestration, provided that the orchestration language supports constructs similar to those from WS-CDL. Even if it is quite a complex language, WS-CDL covers most of the requirements identified for the VISP project (as seen in Table 2).

Considered that that WS-CDL has not reached yet an acceptable maturity level, this technology should be avoided in adopting it for the VISP project. The overall recommendation is to reject, for the moment, WS-CDL, but to monitor its evolution for future convergence with the VISP project needs.

## 4. Orchestration technologies

An orchestration defines (see 0) "*the sequence and conditions in which one Web service invokes other Web services in order to realize some useful function. I.e., an orchestration is the pattern of interactions that a Web service agent must follow in order to achieve its goal.*" While choreography is focused to the composition of services, the orchestration was introduced to describe the flow of communications as a multi-step, long-lived business process from one party's view. Software systems, like workflow engines, are using orchestration languages in order to execute business processes.

A huge effort has been made in the field of orchestration in the last decades. Ten standards have been evaluated for the VISP project. However, the evaluation is concentrated around three main standards, WSBPEL, XPDL and ebXML CPPA. The major technical criteria used in our evaluation, and the corresponding results, are described in *Table 3: **Core technical criteria for orchestration languages***.[1]

*Table 3:* Core technical criteria for orchestration languages

| Technical criteria | BPEL4WS | XPDL | CPPA |
|---|---|---|---|
| Specification of control flow/parallel execution flows | 0 | + | 0/+ |
| Supports substructures like modules/processes | + | + | + |
| Complex structures | 0 | + | + |
| Specification of data flows | + | + | + |
| Compatibility with WSDL/BPMN | +/0 | +/+ | +/0 |
| Specification of bindings / endpoint selection | + | + | + |
| Executable | + | + | + |
| Supports error handling/compensation | +/+ | +/+ | +/0 |
| Supports transaction processing | + | 0 | + |
| Supports roles / human interaction | - | + | + |
| Formal definition | XML | XML | XML Schema |
| Extensibility | 0 | + | + |

### 4.1. ebXML CPPA (Collaboration Protocol Profile and Agreement)

ebXML CPPA is an OASIS standard that provides definitions for the sets of information that are used in business collaborations. Two sets of information are provided: a *Profile* and an *Agreement*. A *Profile* contains data about the technical capabilities of the business partner, needed in order to engage in business collaborations. An *Agreement* contains data that has been agreed to configure the public (or shared) aspects of the protocols to be used in the business collaboration protocols.

ebXML CPPA 2.0 is highly aligned with ebXML BPSS and the ebXML Messaging protocol (also briefly covered in the VISP evaluation of technologies, see 0). It complements several existing standards, including Legacy EDI, XML-based business document standards (such as UBL), and Web Services standards. ebXML CPPA is a mature standard, available as an ISO 15000 standard.

While the CPP part defines the capabilities (technological capabilities and business capabilities, expressed in terms of what business collaborations it supports) of a party to engage in (electronic) business with other parties, the CPA part defines the capabilities that two Parties must agree upon to enable them to engage in (electronic) business for the purposes of the particular CPA. A CPA must be composed from two CPPs.

ebXML CPPA it is best to be used in an e-business project together with ebXML BPSS (for the choreography) and other ebXML related technologies. If UBL is chosen as a business language, ebXML CPPA could be considered for implementing the collaborations between parties. ebXML CPPA offers good support for Web services in the field of business applications, and good support for collaborations and automated negotiation. As an overall conclusion, ebXML CPPA received a weak acceptance from the VISP project.

## 4.2. WSBPEL (Web Services Business Process Execution Language)

WSBPEL was initially developed by BEA, IBM, and Microsoft, over the foundation given by other technologies, like WSFL and XLang. Currently it is under development at OASIS, its current version being released as Committee Draft (WSBPEL 2.0) in December 2005.

WSBPEL is an open, XML based, language. It is intended for the formal specification of business processes and business interaction protocols. WSBPEL defines an interoperable integration model that should facilitate the expansion of automated process integration in both the intra-corporate and the business-to-business space.

WSBPEL supports both the programming-in-the-large and programming-in-the-small paradigms. The programming-in-the-large approach could be used in order to implement the general logic of the process; thus business analysts are able to easily express their requirements, in a formal way, without any involvement in the technical solutions. Next, in the programming-in-the-small approach, the developer could choose appropriate tools in order to specify any architectural problems or programming details.

In the VISP project, WSBPEL was evaluated more as an orchestration language. However, it could be successful used both for choreography and orchestration. The constructions of WSBPEL give to the market/business analysts the ability to describe business processes. Moreover, BPEL, in its very basic version, is abstracted from any technical details, being highly platform-independent.

BPEL4WS 1.1 is recommended for VISP project usage as a low level language, used in order to describe business processes involving both B2B and B2C relations. Since WSBPEL 2.0 is not yet a mature standard, and there is no backward compatibility with BPEL4WS, therefore there is no recommendation for its usage in the VISP project, yet.

## 4.3. XPDL (XML Process Definition Language)

XPDL is a language designed by WfMC, on the basis of WPDL (a previous, non XML, standard from WfMC). Current version of the language is XPDL 2.0, which is backward compatible with previous major releases of this standard.

XPDL is using XML as the mechanism for process definition interchange. XPDL forms a common interchange standard enabling products to continue supporting arbitrary internal representations of the process definition, with corresponding import and export functions to map to (or from) the standard at the product boundary. The XPDL grammar is directly related to these objects and attributes.

An XPDL package has a correspondence with a Business Process Diagram from BPMN. Thus, the XPDL package consists of a set of process definitions, possibly containing references to some subflows, with separate definitions. These subflows are part of the overall process definition. XPDL is highly Web Services oriented, offering seamless integration with Web Services. An activity in a process may invoke a Web Service.

Since XPDL supports most of the requirements of the VISP project, a recommendation of usage was issued for this language, too: XPDL can be used in the VISP project as a low-level workflow implementation language.

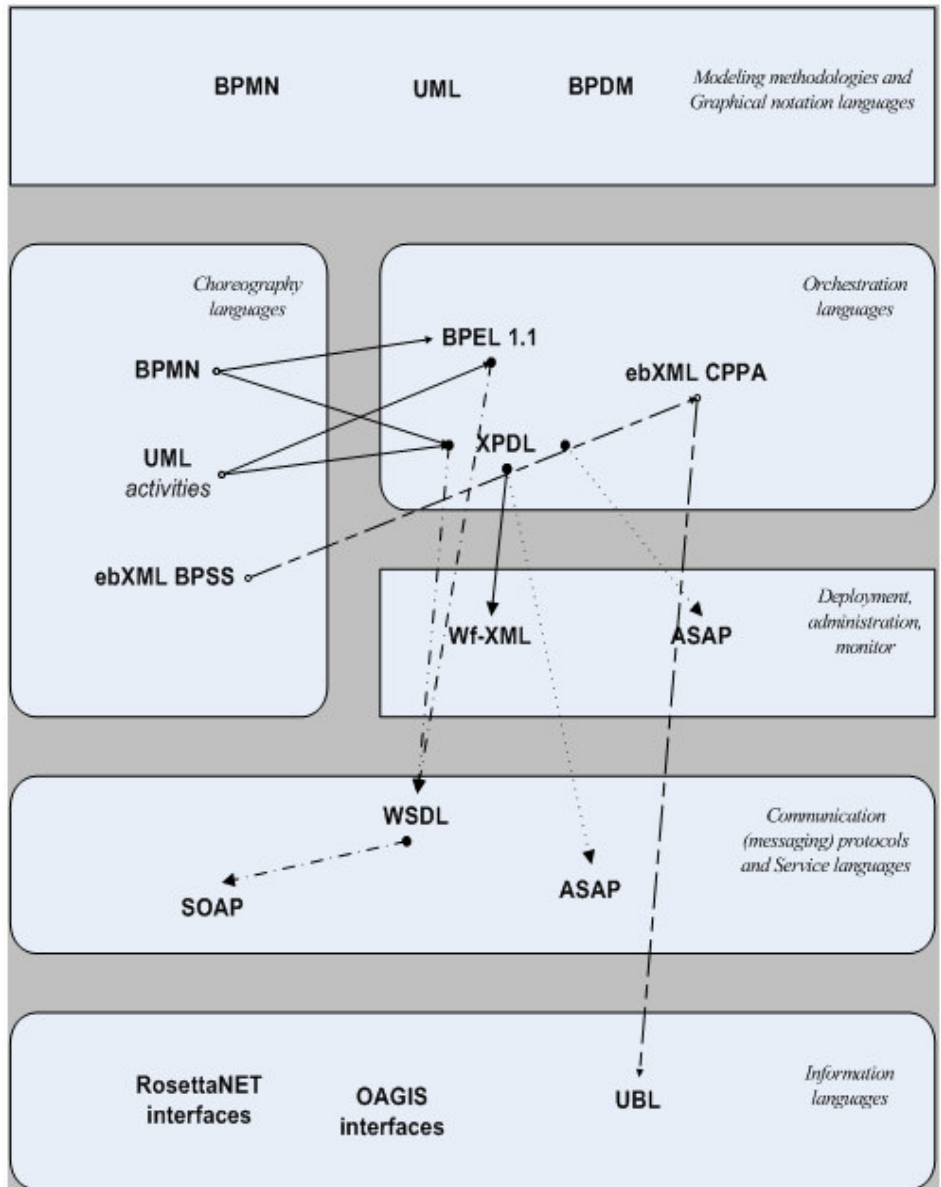

***Fig. 2:*** *VISP stack of standards and some of their relationships*

**Conclusions**

The analysis of target technologies in the fields of business process management and workflows was conducted in three main directions: market penetration, technical information, and applicability to the VISP project. For each analyzed technology some overall usage recommendations were issued, considering their strengths and weaknesses that were emphasized from the three research directions. The results of the analysis have revealed the VISP stack of technologies (as summarized in *Fig. 2: VISP stack of standards and some of their relationships*) to be used in the development of the VISP project.

**Acknowledgements**

This work has been carried out with partial funding from the EU through the IST project VISP, IST-FP6-027178. Further information and current development status on the VISP project are available on the VISP site, http://www.visp-project.org.

**References**

[A+03]   W.M.P. van der Aalst, A.H.M. ter Hofstede, B. Kiepuszewski, A.P. Barros - *Workflow Patterns*, *Distributed and Parallel Databases*, 14 (3), pp. 5—51, July 2003

[A+03]   T. Andrews et all - *Business Process Execution Language for Web Services*, Version 1.1, 5 May 2003.

[OAS05]  OASIS - *Web Services Business Process Execution Language*, Version 2.0, Committee Draft, December 2005 (on OASIS web site, http://www.oasis-open.org/)

[OAS06]  OASIS - *ebXML Business Process Specification Scheme*, v2.0.2, January 2006 (on OASIS web site, http://www.oasis-open.org/)

[VISP06] VISP Project FP6-IST-2004-027178, D2.1 - *VISP Workflow Technologies Functional Analysis and Comparison*, Deliverable D2.1, March 2006 (on the VISP Project web site, http://www.visp-project.org)

[WfM02]  WfMC - *Workflow Process Definition Interface — XML Process Definition Language (XPDL)*, Document WFMC-TC-1025,

Version 1.0, October 2002 (on WfMC web site, http://www.wfmc.org/)

[WfM05] WfMC - *Process Definition Interface — XML Process Definition Language*, Document WFMC-TC-1025, Version 2.00, October 2005 (on WfMC web site, http://www.wfmc.org/)

[Whi04] S.A. White - *Processing Modeling Notations and Workflow Patterns*, *Business Process Trends*, March 2004

[W3C] Glossary — Dictionary, http://www.w3.org/2003/glossary/

[W3C05] W3C - *Web Services Choreography Description Language; Version 1.0*, W3C Candidate Recommendation, November 2005 (on W3C web site, http://www.w3c.org/)

---

[1] A standard satisfying the criterion is marked with +; if the criterion is partially satisfied the 0 sign is used; and – denotes a standard that does not satisfy the criterion.